\documentclass[manuscript]{aastex}

\usepackage{graphicx}
\usepackage[space]{grffile}
\usepackage{latexsym}
\usepackage{textcomp}
\usepackage{longtable}
\usepackage{multirow,booktabs}
\usepackage{amsfonts,amsmath,amssymb}
\usepackage{url}
\usepackage{hyperref}
\hypersetup{colorlinks=false,pdfborder={0 0 0}}
\usepackage{graphicx}
\usepackage[space]{grffile}
\usepackage{url}
\usepackage[utf8]{inputenc}
\usepackage{hyperref}
\usepackage{longtable}
\usepackage{natbib}

\hypersetup{colorlinks=false,pdfborder={0 0 0}}
\shorttitle{CS emission near MIR-bubbles}
\shortauthors{C. Watson}

\shorttitle{CS emission near MIR-bubbles}
\shortauthors{C. Watson et al.}

\begin{document}


\title{CS emission near MIR-bubbles}


\author{C. Watson}
\affil{Manchester University, Dept. of Physics, 604 E. College Ave., North Manchester, IN 46962}
\email{cwatson@manchester.edu}

\author{Kathryn Devine}
\affil{College of Idaho, Dept. of Physics, 2112 Cleveland Blvd, Caldwell, ID 83605}
\email{KDevine@collegeofidaho.edu}
 
\author{N. Quintanar}
\affil{Texas A\&M University, Dept. of Nuclear Engineering, 401 Joe Routt Blvd, College Station, TX 77843}
\email{nrquintanar@tamu.edu}

\author{T. Candelaria}
\affil{New Mexico Institute of Mining and Technology, Dept. of Physics, 801 Leroy Place, Socorro, NM 87801}
\email{tcandela@nmt.edu}

\begin{abstract}
We survey 44 young stellar objects located near the edges of mid-IR-identified bubbles in CS (1-0) using the Green Bank Telescope. We detect emission in 18 sources, indicating young protostars that are good candidates for being triggered by the expansion of the bubble. We calculate CS column densities and abundances. Three sources show evidence of infall through non-Gaussian line-shapes. Two of these sources are associated with dark clouds and are promising candidates for further exploration of potential triggered star formation. We obtained on-the-fly maps in CS (1-0) of three sources, showing evidence of significant interactions between the sources and the surrounding environment.%
\end{abstract}%


\bibliographystyle{apj}

\section{Introduction}
Prior to post-main-sequence evolution, ionizing radiation is one of the most important mechanisms by which massive stars influence their surrounding environments. This ionizing radiation may potentially trigger subsequent star-formation. The influence of ionizing radiation is observed in the form of bubble-shaped emission in the 8 $\mu$m band of the Spitzer-GLIMPSE survey of the Galactic Plane \citep{Benjamin2003}. \citet{Churchwell2006,Churchwell2007} observed bubble-shaped 8 $\mu$m emission to be common throughout the Galactic plane. \citet{Watson2008, Watson2009} found 24 $\mu$m and 20 cm emission centered within the 8 $\mu$m emission and interpreted the bubbles seen in the GLIMPSE data as caused by hot stars ionizing their surroundings, creating 20 cm free-free emission, and at larger distances exciting PAHs, creating 8 $\mu$m emission. \citet{Deharveng2010} also interpreted the bubbles as classical HII regions.

\citet{Watson2010} used 2MASS and GLIMPSE photometry and Spectral Energy Distribution (SED)-fitting to analyze the YSO population around 46 bubbles and found about a quarter showed an overabundance of YSOs near the boundary between the ionized interior and molecular exterior. These YSOs are candidates for being triggered by the expanding ionization and shock fronts created by the hot star. Star formation triggered by previous generations of stars is known to occur but the specific physical mechanism is still undetermined. The collect-and-collapse model \citep{Elmegreen1977} describes ambient material swept up by the shock fronts which eventually becomes gravitationally unstable, resulting in collapse. Other mechanisms, however, have been proposed. Radiatively-driven implosion \citep{1994A&A...289..559L}, for example, describes clumps already present in the ambient material whose contraction is aided by the external radiation of the hot star.

Bubbles with an overabundance of YSOs along the bubble-interstellar medium (ISM) boundary are a potentially excellent set of sources to study the mechanisms of triggered star-formation. The method of identifying YSOs through photometry, however, is limited. \citet{Robitaille_2006} showed that YSO age is degenerate with the observer's inclination angle. An early-stage YSO and a late-stage YSO seen edge on, so the accretion or debris disk is observed as thick and blocking the inner regions, can appear similar, even in the IR. Thus, we require other diagnostics of the YSOs along the bubble edge to determine the youngest, and most likely to have been triggered, YSOs. Additionally, a line-diagnostic allows us to rule out any line-of-sight coincidence associations.

For the current project we selected a subset of the bubbles identified above to identify those YSOs associated with infall, outflows or hot cores by observing the CS (1-0) transition near 49 GHz with the Green Bank Telescope (GBT\footnote{The National Radio Astronomy Observatory is a facility of the National Science Foundation operated under cooperative agreement by Associated Universities, Inc.}). CS is a probe of young star-formation. It has been detected in outflows from protostars, infall, disks and in hot cores \citep{1997A&A...317L..55D,1996A&AS..115...81B,Morata2012}. The chemistry is, naturally, complex, and it appears that CS can play several roles \citep{Beuther2002}, such as tracing outflows \citep{Wolf-Chase1998} or hot cores \citep{1997MNRAS.287..445C}. Our aim here is to use CS as a broad identifier of young star-formation and use any non-Gaussian line-shapes to infer molecular gas behavior.

After describing the CS survey and CS mapping observations ($\S$ 2) and numerical results ($\S$ 3), we analyze the Herschel-HiGAL emission toward all our sources to determine, along with our CS detections, the CS abundances ($\S$ 4.1). We also analyze three sources with evidence of rapid infall ($\S$ 4.2). We end with a summary of the conclusions.

\section{Observations}
Candidate YSO locations were identified using the SED fitter tool developed by \citet{Robitaille_2006, Robitaille_2007}. Briefly, this method uses the 2MASS \citep{1994ExA.....3...65K} and GLIMPSE point source catalogues to identify sources that are not well-fit by main-sequence SEDs and are well-fit by YSO SEDs. \citet{Watson2010} fit all point sources within 1\arcmin of the bubble edges using this method. From this set of point sources, four sources were selected near each bubble based on association with either diffuse, bright 8 $\mu$m emission or IR dark clouds. Forty point sources in total were selected. The names, Galactic longitude and Galactic latitude are reported in Table 2. Each point source was observed for CS using the Green Bank Telescope (GBT) for two 5 minute integrations. The spectrometer was set-up in frequency switching mode to maximize on-source observing time. The setup parameters and calibration sources are listed in Table 1.

\begin{deluxetable}
{rr}
\label{configuration}
\tablecaption{Observing Parameters}
\startdata
Bandwidth   &50 MHz\\
Channel width   &1.5 kHz\\
Rest frequency  &48.99095 GHz\\
Frequency switching shift   &8 MHz\\
Pointing calibration    &1751+0930\\
    &1850-0001\\
    &2025+3343\\
Flux calibration    &NGC7027\\
\enddata

\end{deluxetable}

Data were calibrated and analyzed using GBTIDL. Typical system temperatures were between 105 K and 120 K. Typical rms noise in the resulting calibrated spectra was 0.20 K. Non-detections and detections are listed in Tables 2 and 3, respectively. We estimate uncertainty due to flux calibration of 20\%.

In addition to single pointings, we mapped three regions (N56, N65 2 and N77 1) that displayed strong CS emission. The map sizes were 1\arcmin x1 \arcmin (N56 and N77-1) and 2\arcmin x2\arcmin (N65-2), both using a Nyquist-sampling step-size of 6.12\arcsec.

Observations were used from the Hi-Gal \citep{Molinari2010} project, a Herschel Space Telescope imaging survey of the Galactic plane. This survey observed all the sources in this study at wavelengths between 60 $\mu$m and 600 $\mu$m. Data were downloaded from the Spitzer Science Center website. Level 2 data products were used, which have been fully calibrated.

\begin{deluxetable}
{rrr}
\tablecaption{CS Non-Detections}
\tablehead{\colhead{Name} &\colhead{l($^\circ$)} &\colhead{b($^\circ$)}}
\startdata
N62-2	&34.329	&0.195	\\
N62-3	&34.317	&0.197	\\
N65-3	&34.963	&0.310	\\
N65-4	&35.049	&0.330	\\
N77-3	&40.407	&-0.037	\\
N77-4	&40.409	&-0.033	\\
N82-1	&42.122	&-0.635	\\
N82-2	&42.128	&-0.636	\\
N82-3	&42.114	&-0.616	\\
N82-4	&42.112	&-0.658	\\
N90-3	&43.748	&0.0754	\\
N90-4	&43.735	&0.0629	\\
N92-1	&44.359	&-0.825	\\
N92-4	&44.335	&-0.824	\\
N117-1	&54.102	&-0.094	\\
N117-2	&54.076	&-0.085	\\
N123-1	&57.562	&-0.297	\\
N123-3	&57.567	&-0.285	\\
N123-4	&57.564	&-0.280\\	
N128-1	&61.688	&0.990	\\
N128-2	&61.703	&0.988	\\
N128-3	&61.625	&0.953	\\
N128-4	&61.704	&0.921	\\
\enddata

\end{deluxetable}

\section{Results}
\subsection{CS Point Sources}
\begin{figure}[h!]
\begin{center}
\includegraphics[width=0.98\columnwidth]{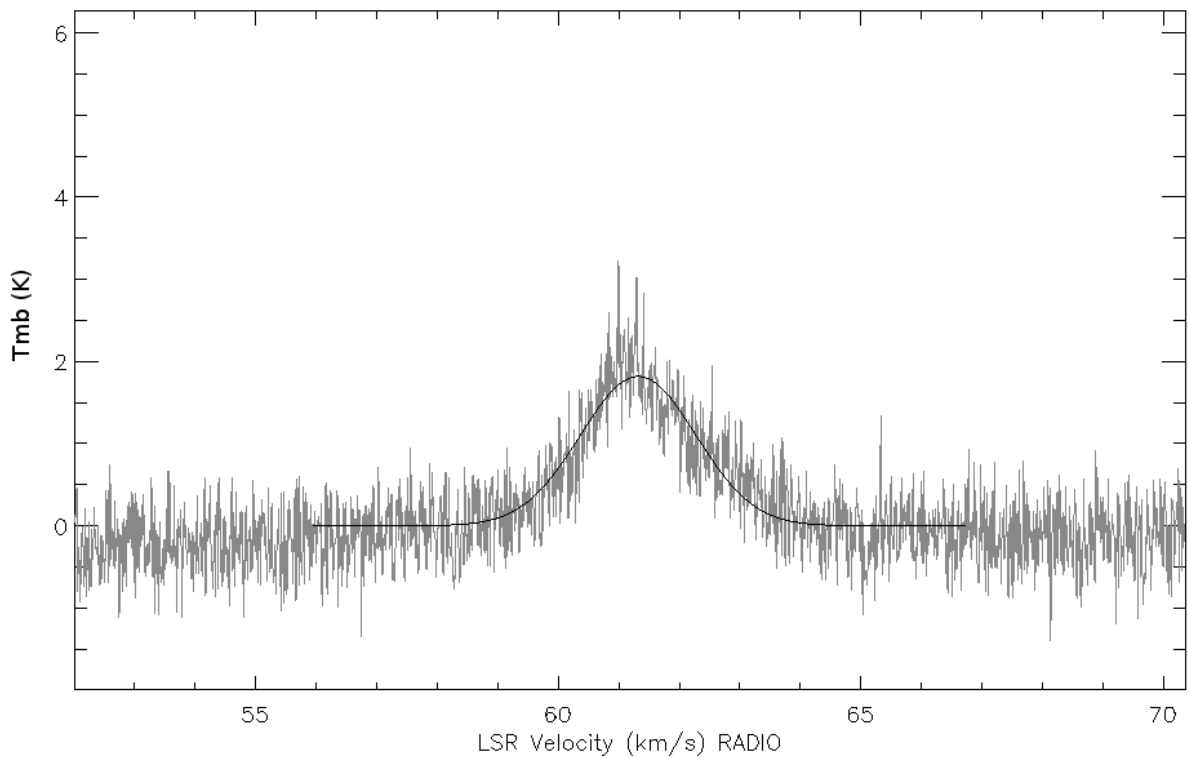}
\caption{Detection of CS emission (grey) toward a candidate YSO (N92-3) on the rim of bubble N92. The Gaussian fit to the emission is shown in black. \label{N92spectrum}%
}
\end{center}
\end{figure}

Eighteen sources displayed emission greater than 3$\sigma$. A typical spectrum is shown in Figure \ref{N92spectrum}. Emission lines were fit using fitgauss, the standard Gaussian fitting routine in GBTIDL. Fitting parameters (amplitude in T$_{mb}$ units, central velocity and FWHM) are listed in Table \ref{fitting}. For sources that displayed a double peak, two simultaneous Gaussian functions were fit to the emission and are listed in consecutive rows. CS column densities, \begin{math} N_{CS} \end{math}, were calculated assuming LTE, optically thin emission and an excitation temperature T$_{ex}$=15 K, a typical ISM value (see review in \citet{Zinnecker2007}). Increasing or decreasing the assumed excitation temperature by 5 K changes the column density by about 30\%. If CS(1-0) is optically thick, as we assume for three sources in section 4.2 below, then our calculation would be a lower limit. Given these assumptions we used the following relation (see \citet{Miettinen2012} for a detailed discussion of the relations below):

\begin{equation*}
N_{CS} = \frac{3 k_B \epsilon_0}{2 \pi^2}\frac{1}{\nu \mu^2_{el}S}\frac{Z_{rot}(T_{ex})}{g_K g_I}\frac{e^{E_u/k_B T_{ex}}}{1-\frac{F(T_{bg})}{F(T_{ex})}}\int T_{MB}dv
\end{equation*}
where

\begin{eqnarray*}
g_K= g_I &= 1\nonumber\\
\mu^2_{el} S &= 3.8 \;\mathrm{Debye}^2\nonumber\\
Z_{rot} &= {\bf 0.8556}\; T_{ex}-0.10\nonumber\\
F(T) &= \frac{1}{e^{h\nu/k_BT}-1}.\nonumber\\
\end{eqnarray*}
Here $\epsilon_0$ is the vacuum permittivity, $\mu_{el}$ is the permanent electric dipole moment, S is the line strength, Z$_{rot}$ is the rotational partition function, $\nu$ is the frequency, g$_K$ is the K-level degeneracy, g$_I$ is the reduced nuclear spin degeneracy, E$_u$ is the energy of the upper-transition state, T$_{ex}$ is the excitation temperature and T$_{bg}$ is the background temperature. The dipole moment line strength ($\mu^2_{el}$ S) is taken from the JPL spectral line catalog \citep{Pickett1998}. The partition function (Z$_{rot}$) is a linear fit to JPL data between T=37 to 75 K. T$_{bg}$ was taken to be the cosmic microwave background temperature, 2.725 K. The uncertainty in the fit amplitudes and derived column densities is dominated by our flux-calibration uncertainty. Since the relationships are linear, we estimate the uncertainty in both as 20\%.

\begin{deluxetable}
{lllllll}
\label{fitting}
\tablecaption{Gaussian fitting parameters for CS detections.}
\tablehead{
\colhead{Name}    &\colhead{l ($^\circ$)}  &\colhead{b ($^\circ$)}  &\colhead{T$_{mb}$ (K)}  &\colhead{Vel$_{LSR}$ (km/s)} &\colhead{FWHM (km/s)}    &\colhead{N$_{CS}$ (cm$^{-2}$)}}
\startdata
N62-1	&	34.352	&	0.192	&	1.3	&	57.6	&	0.9	&	3.4$\times$10$^{14}$	\\
	&		&		&	3.5	&	56.4	&	1.2	&	1.3$\times$10$^{15}$	\\
N62-2	&	34.329	&	0.195	&	1.1	&	57.2	&	1.9	&	6.1$\times$10$^{14}$	\\
N65-1	&	35.044	&	0.327	&	1.6	&	51.3	&	1.7	&	8.1$\times$10$^{14}$	\\
N65-2	&	35.025	&	0.350	&	2.4	&	50.4	&	2.2	&	1.6$\times$10$^{15}$	\\
	&		&		&	10.9	&	53.3	&	4.1	&	1.3$\times$10$^{16}$\\
N65-4	&	35.049	&	0.330	&	2.2	&	51.3	&	1.9	&	1.1$\times$10$^{15}$	\\
N77-1	&	40.437	&	-0.044	&	3.1	&	68.0	&	1.6	&	1.4$\times$10$^{15}$	\\
N77-2	&	40.422	&	-0.024	&	4.1	&	69.5	&	2.6	&	2.9$\times$10$^{15}$	\\
N82-5	&	42.125	&	-0.623	&	5.7	&	66.4	&	1.8	&	2.9$\times$10$^{15}$	\\
N90-1	&	43.788	&	0.083	&	1.2	&	35.2	&	0.7	&	2.1$\times$10$^{14}$	\\
N90-2	&	43.792	&	0.089	&	1.0	&	36.0	&	0.6	&	1.4$\times$10$^{14}$	\\
	&		&		&	2.6	&	35.4	&	0.6	&	3.9$\times$10$^{14}$	\\
N92-2	&	44.349	&	-0.803	&	1.3	&	61.6	&	1.4	&	5.8$\times$10$^{14}$	\\
N92-3	&	44.334	&	-0.818	&	1.9	&	61.3	&	2.6	&	1.5$\times$10$^{15}$	\\
N117-3	&	54.107	&	-0.044	&	2.3	&	40.9	&	1.7	&	1.2$\times$10$^{15}$	\\
	    &		    &		    &	3.8	&	38.4	&	2.3	&	2.7$\times$10$^{15}$	\\
N123-2	&	57.578	&	-0.284	&	2.3	&	-9.3	&	1.9	&	1.4$\times$10$^{15}$	\\
N133-1	&	63.125	&	0.442	&	1.0	&	20.7	&	2.2	&	7.0$\times$10$^{14}$	\\
N133-2	&	63.132	&	0.415	&	1.4	&	19.3	&	2.2	&	7.3$\times$10$^{14}$	\\
N133-3	&	63.179	&	0.440	&	1.4	&	23.3	&	2.2	&	7.4$\times$10$^{14}$	\\
N133-4	&	63.152	&	0.441	&	1.5	&	24.0	&	4.6	&	1.7$\times$10$^{15}$	\\
	    &		    &		    &	3.2	&	23.3	&	1.8	&	1.4$\times$10$^{15}$	\\
\enddata

\end{deluxetable}

\subsection{CS Mapping}
\begin{figure}[h!]
\begin{center}
\includegraphics[width=0.98\columnwidth]{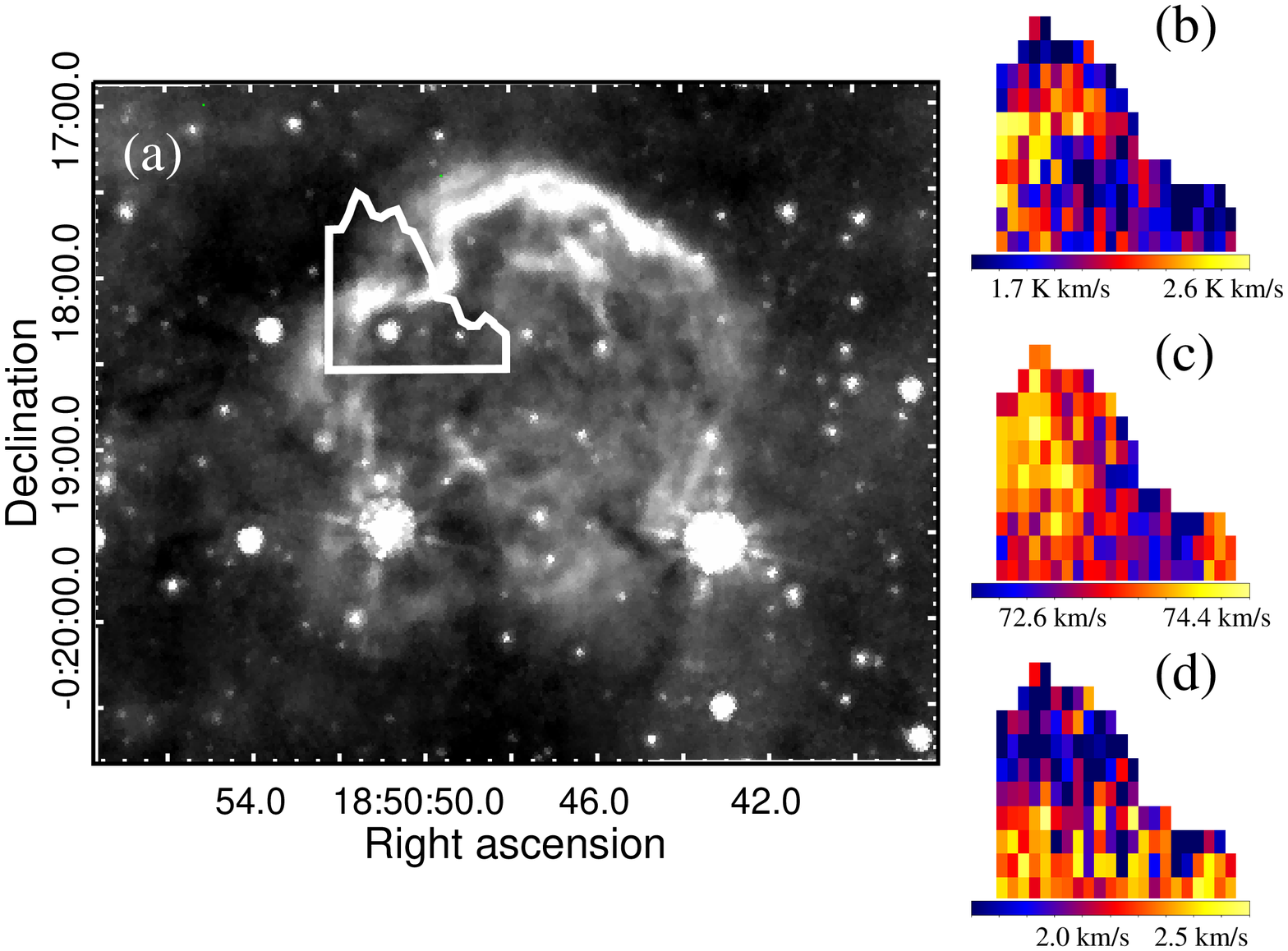}
\caption{N56: a) The region surrounding N56 mapped in CS (white outline) overlaid on a GLIMPSE 8 $\mu$m survey image. b) the integrated intensity map c) the average velocity map d) the velocity width map.
\label{N56map}%
}
\end{center}
\end{figure}

\begin{figure}[h!]
\begin{center}
\includegraphics[width=0.98\columnwidth]{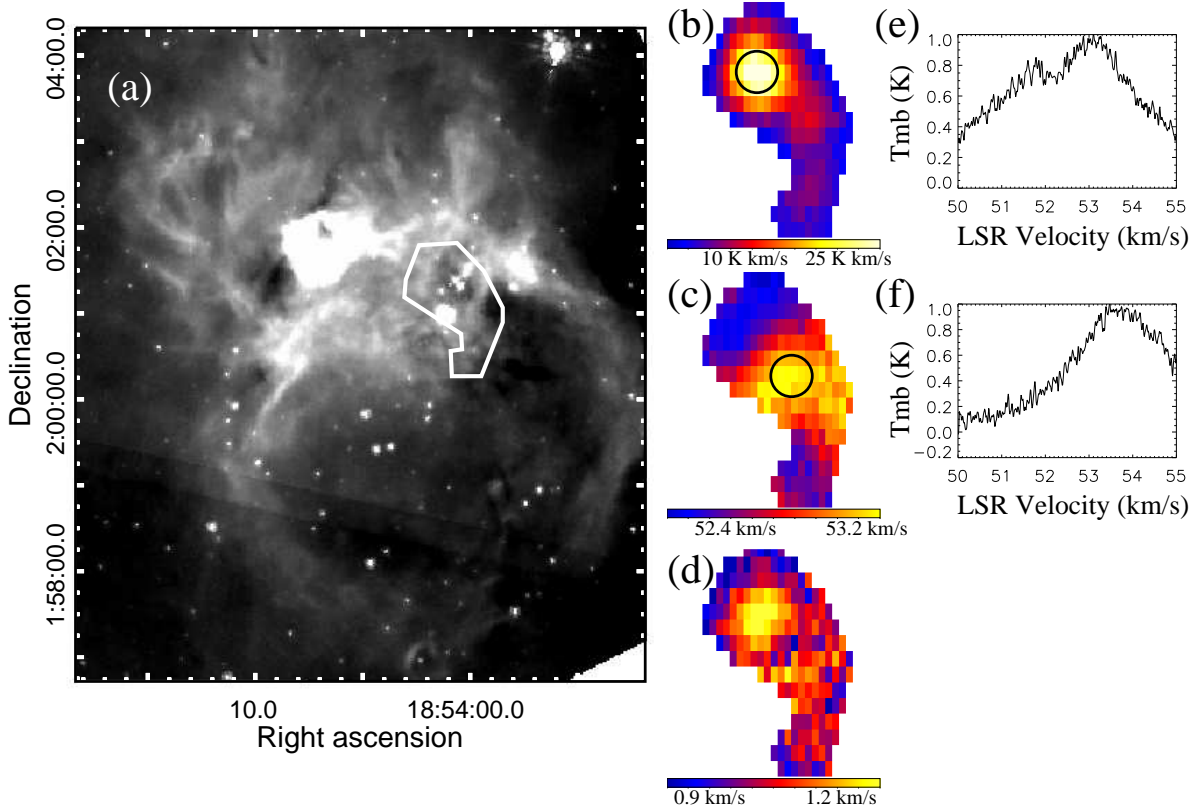}
\caption{N65: a) The region surrounding N65 2 mapped in CS (white outline) overlaid on a GLIMPSE 8 $\mu$m survey image b) the integrated intensity map c) the average velocity map d) the velocity width map, e) a spectrum centered on the primary peak marked in {\bf part b}, f) a spectrum centered on the secondary peak and the shift in velocity, marked in {\bf part c}.
\label{N65map}
}
\end{center}
\end{figure}

\begin{figure}[h!]
\begin{center}
\includegraphics[width=0.98\columnwidth]{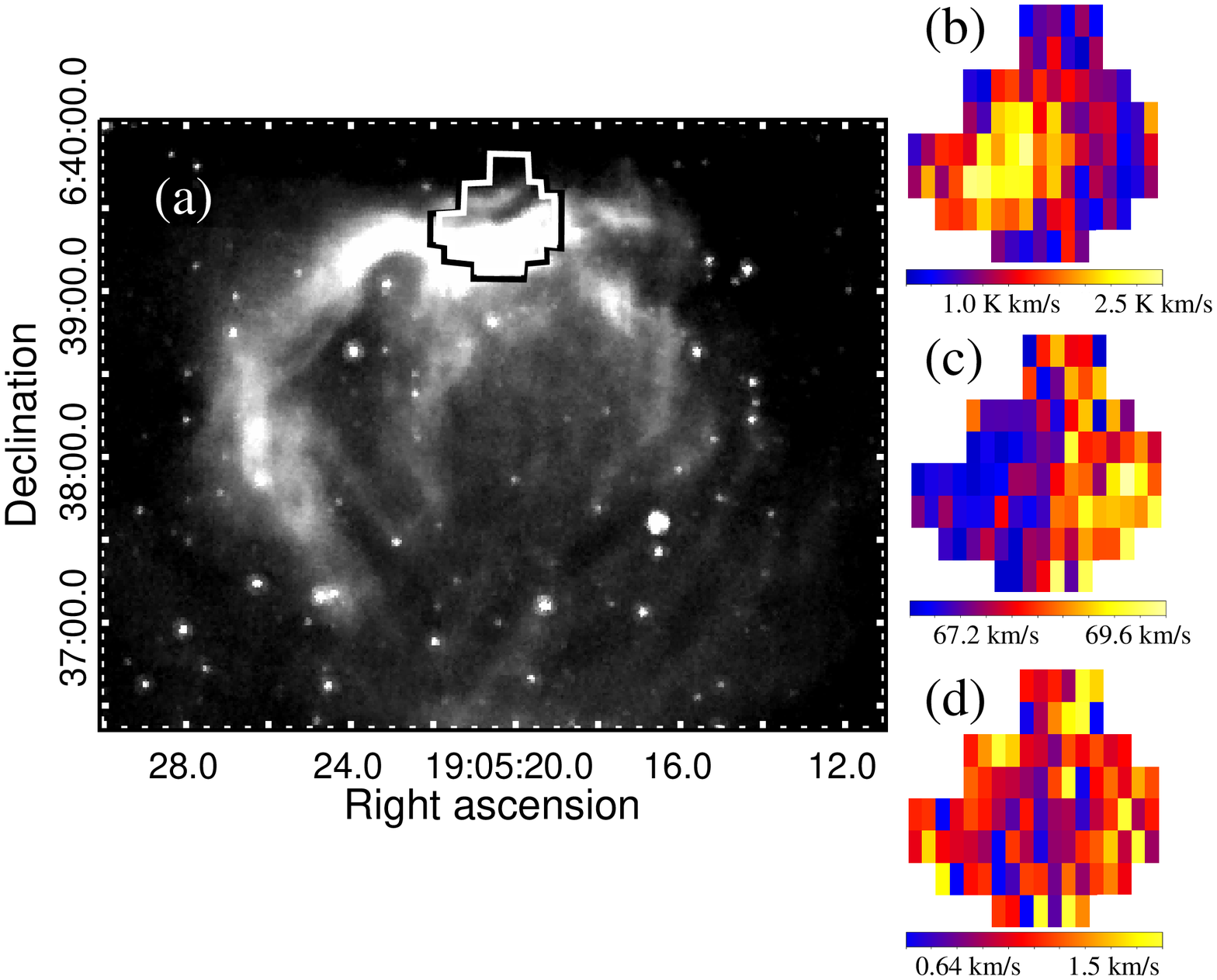}
\caption{N77: a) The region surrounding N77 1 mapped in CS (white and black outline) overlaid on a GLIMPSE 8 $\mu$m survey image b) the integrated intensity map c) the average velocity map d) the velocity width map.
\label{N77map}
}
\end{center}
\end{figure}

The three mapped regions are shown as boxes overlaid on the GLIMPSE 8 $\mu$m images in Figures \ref{N56map}-\ref{N77map}. The maps were cropped to just those regions showing emission above 3$\sigma$. The regions containing emission were then exported to a FITS file using custom-built IDL tools and the FITS-format datacubes were analyzed using CASA. The standard moment maps (total intensity, average velocity and velocity width) are shown next to the corresponding 8 $\mu$m emission images in Figures \ref{N56map}-\ref{N77map}.

The three maps of CS in N56, N65, and N77 (Figs. \ref{N56map}-\ref{N77map}) show evidence of supersonic gas motion in areas near each YSO. N56 has a weak peak in CS that corresponds to a velocity shift of $\sim$2 km/s in the gas immediately to the north. The CS emission in N65 has a clear, strong comma-like shape. The peak in emission is clearly near the top of the shape, but there is a slight, secondary peak below and to the right in Fig 3b. There is also a shift in gas velocity of $\sim$1 km/s at the same location. Two spectra from the map, centered at the primary and secondary peaks, are shown in Figs. 3e and f. The spectrum at the primary peak shows two components, with the red-shifted component stronger. At the secondary peak, this component appears to shift further redward. We interpret the secondary peak, shift in first-moment map and the double-Gaussian peak from single pointing spectra as all caused by at least two overlapping clouds at different velocities. Using this interpretation, the emission peak (Fig. 3b) is highest where emission from both clouds is present.  This peak is coincident with the blue-shifted gas (Fig. 3c) because emission from the bluer cloud is present at the location of the pointing shown in Fig. 3b but not at the location of the pointing shown in Fig. 3c. There may be further motion within each cloud, which could be responsible for the redward shift of the stronger component. The least distinct of the regions, N77 shows weak emission with some evidence of shifts in velocity of $\sim$1 km/s. These shifts could be caused by several mechanics: outflow from or infall toward the YSO or shock-induced velocity shifts caused by the expanding HII region. The limited nature of the data prevents an exclusive interpretation.

\section{Analysis}

\subsection{CS Abundance}
To calculate the abundance of CS, we first must estimate the total gas column density along each line of sight. We used FIR (60$\mu$m-600$\mu$m) imaging taken as part of the HiGal survey. Within CASA, we measured the integrated emission in all five survey bands in regions exactly coincident with the GBT beamsize, centered at each source of CS emission. The emission was then modeled as a modified blackbody:

\begin{equation*}
B_{mod} =B_0 \frac{2 h \nu^{3+\beta}}{c^3}\frac{1}{e^{\frac{h\nu}{kT}}-1}
\end{equation*}

where T is temperature, $\nu$ is frequency, B$_0$ is a scaling constant and $\beta$, the emissivity index, is assumed to be 2 \citep{Desert2008}. B$_0$ and T were taken as free parameters and a Levenberg{\textendash}Marquardt algorithm was used to find the best-fit. The fitting was done to the flux density in Jy, so B$_0$ carries these units. The total column density, N$_{tot}$, was calculated following \citet{Miettinen_2010}. Briefly, we used the following relations:

\begin{eqnarray*}
N_{tot} = \frac{I}{B_\nu \mu m_H \kappa R_d}\nonumber\\
B_\nu = \frac{2 h \nu^3}{c^2 (e^\frac{h \nu}{kT}-1)}\nonumber\\
I = 3.73\times 10^{-16} B_{mod} \left(\frac{1"}{\theta}\right)^2\nonumber\\
\kappa = \kappa_{1.3mm} \left(\frac{\lambda}{1.3mm}\right)^{-\beta}\nonumber\\
\end{eqnarray*}

where m$_H$ is the mass of hydrogen and 3.73 x 10$^{-16}$ converts the surface brightness from Jy/(1\arcsec beam) to SI units. We make the following assumptions: the opacity at 1.3 mm is $\kappa_{1.3mm}$ = 0.11 $\frac{m^2}{kg}$, appropriate for ice-covered dust grains from \citet{1994A&A...291..943O}, $\theta$=15.0\arcsec , the beamsize of the GBT at 49 GHz, the mean molecular weight $\mu$ = 2.3 and dust to mass ratio $R_d$ = 1/100. Note that B$_\nu$, B$_{mod}$ and $\kappa$ all require a choice of frequency or wavelength. However, these dependencies cancel in the final calculation of N$_{tot}$. These results are summarized in Table \ref{mbb}, where we report the flux density at five wavelength bands for each CS detection, the best-fit temperature, the total column density and the CS abundance. We estimate the error in determining the extended flux to be dominated by defining the edge of the object. These sources all appear extended in the Herschel bands and some lie in confused regions. Thus, the gas sampled by FIR and CS are likely different. This difference should lead to a cautious association between the dust temperatures and the CS emission. To estimate the influence of this uncertainty on the calculated properties, we examined the effect of a 20\% change up or down in FIR flux. The results were a change of 4 K in temperature and 20\% in column density.

For those sources where the modified blackbody model was a poor fit, as judged by eye, we have excluded the temperature, column density and abundance. The cause for the poor fit in these cases appeared to be caused by emission extending well outside the the GBT beam. For these poorly-fit sources, the fluxes reported here probably do not represent the emission from the same object. For those sources with a double-Gaussian CS line profile, we add the CS column densities calculated using both Gaussians. If this shape is caused by optical depth effects, as we discuss below, than the reported column density would be a lower limit.

\begin{deluxetable}
{rrrrrrrrr}
\label{mbb}
\rotate
\tabletypesize{\small}
\tablecaption{Modified blackbody fitting of Herschel HiGal observations toward CS-detections.}
\tablehead{
\colhead{Name}	&\colhead{Blue}	&\colhead{Red}	&\colhead{PSW}	&\colhead{PMW}	&\colhead{PLW}	&\colhead{Temp.}	&\colhead{N$_{Tot}$}	&\colhead{CS Abundance }\\
& \colhead{60-85$\mu$m} &\colhead{130-210$\mu$m} &\colhead{250$\mu$m} &\colhead{350$\mu$m} &\colhead{500$\mu$m}\\
\colhead{} &\colhead{(Jy)}	&\colhead{(Jy)}	&\colhead{(Jy)}	&\colhead{(Jy)}	&\colhead{(Jy)}	&\colhead{(K)}	&\colhead{($\times$10$^{21}$ cm$^{-2}$)}	&\colhead{($\times$10$^{-7}$)}}
\startdata
N62-1	&	31.1	&	106.4	&	157.4	&	78.0	&	30.6	&	---	&	---	&	---	\\
N62-2	&	86.1	&	203.7	&	303.9	&	142.7	&	56.1	&	---	&	---	&	---	\\
N65-1	&	29.1	&	19.2	&	13.1	&	3.2	&	2.1	&	30	&	2.92	&	2.76	\\
N65-2	&	831.7	&	553.7	&	97.4	&	19.4	&	7.7	&	32	&	15.8	&	9.36	\\
N65-4	&	9.9	&	13.4	&	9.4	&	3.0	&	2.1	&	24	&	4.10	&	2.79	\\
N77-1	&	9.8	&	12.7	&	7.2	&	1.5	&	0.5	&	25	&	1.93	&	7.32	\\
N77-2	&	1.5	&	6.9	&	13.0	&	5.4	&	2.8	&	---	&	---	&	---	\\
N82-5	&	1065.3	&	797.1	&	416.1	&	173.3	&	62.9	&	29	&	165	&	0.17	\\
N90-1	&	91.6	&	155.3	&	276.7	&	133.9	&	51.3	&	---	&	---	&	---	\\
N90-2	&	5.6	&	12.3	&	17.6	&	7.8	&	3.1	&	---	&	---	&	---	\\
N92-2	&	5.8	&	7.4	&	4.7	&	2.2	&	0.8	&	25	&	2.83	&	2.05	\\
N92-3	&	4.9	&	8.0	&	4.8	&	3.1	&	2.1	&	23	&	4.61	&	3.29	\\
N117-3	&	9.1	&	20.1	&	9.8	&	5.5	&	1.9	&	23	&	8.85	&	4.42	\\
N123-2	&	4.6	&	7.7	&	6.5	&	2.8	&	1.0	&	22	&	4.64	&	2.91	\\
N133-1	&	28.5	&	16.0	&	4.8	&	4.4	&	0.9	&	32	&	3.46	&	2.01	\\
N133-2	&	5.2	&	7.8	&	3.5	&	1.1	&	0.8	&	25	&	1.38	&	5.28	\\
N133-3	&	13.8	&	12.1	&	5.0	&	3.0	&	1.1	&	28	&	3.19	&	2.31	\\
N133-4	&	21.2	&	24.7	&	13.2	&	5.5	&	1.8	&	26	&	6.47	&	4.68	\\
\enddata

\end{deluxetable}

\begin{figure}[h!]
\begin{center}
\includegraphics[width=0.98\columnwidth]{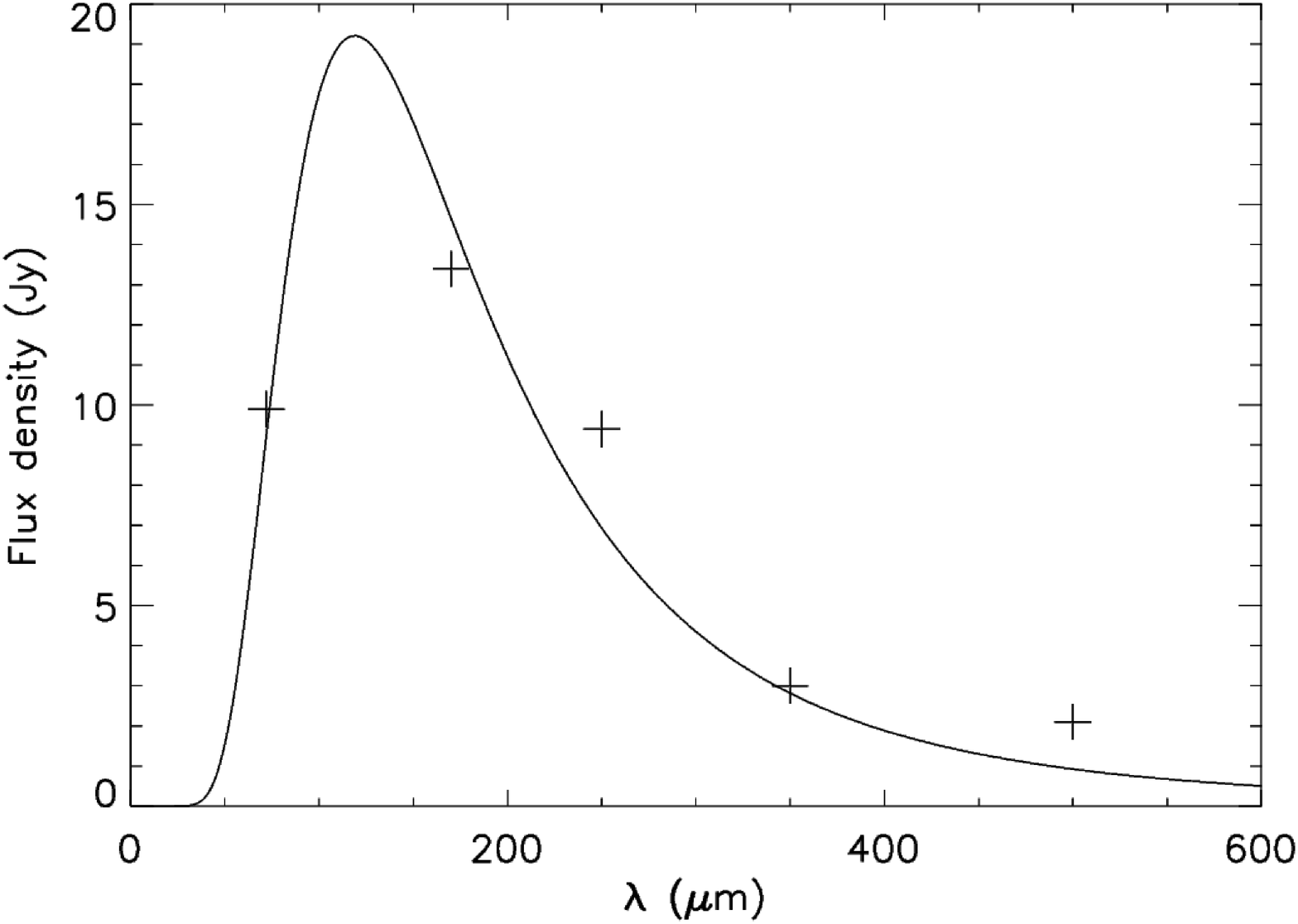}
\caption{N65 4: Mid-IR Spectral Energy Distribution (crosses), obtained by integrating images from HiGal, a Herschel survey of the Galactic plane, in a region equivalent to the GBT beam centered at the pointing location for N65-4 given in Table 3. A modified blackbody model (line) using $\beta$=2 was fit to the data. The model was used to calculate temperature (24 K) and column density (4.1 $\times$ 10$^{21}$ cm$^{-2}$). \label{N65bb}%
}
\end{center}
\end{figure}

\subsection{Infall}
\begin{figure}[h!]
\begin{center}
\includegraphics[width=0.98\columnwidth]{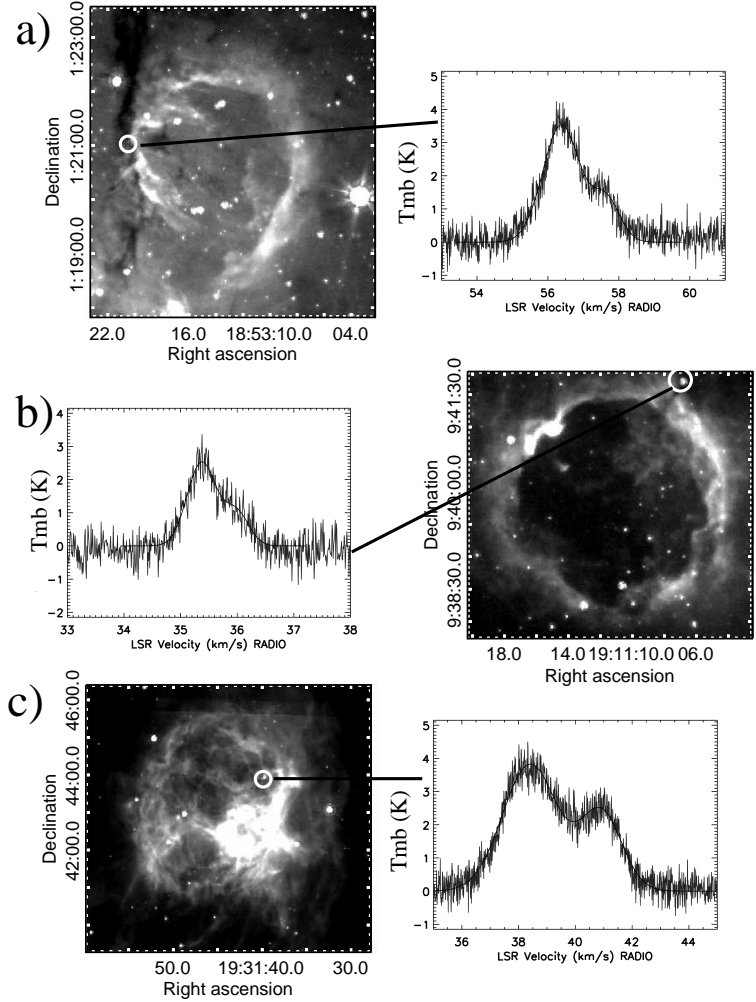}
\caption{Non-Gaussian line-profiles of CS emission toward three sources. The thin black line represents the observed spectrum; the thick black line represents the double-Gaussian fit to the data. The images are 8 $\mu$m emission taken from the GLIMPSE/Spitzer survey. The solid white circle indicates the position of the YSO and the size of the GBT beam. Sources are a) N62-1, b) N90-2 and c) N117-3. \label{infall}%
}
\end{center}
\end{figure}

Four sources, N62-1, N65-2, N90-2 and N117-3, have a non-Gaussian line profile. In three cases the line profile is stronger on the blue-side (see Fig. \ref{infall}). Of these three sources, N117-3 has the strongest red-shifted emission, with two clear peaks present. The line-profiles of N62-1 and N90-2 are single-peaked but with a plateau on the red-shifted side. We interpret these three profiles as evidence of infall. N62-1 and N90-2 both are located in infrared dark clouds that intersect their nearby bubble (N62 and N90). Thus, infall, if present, could be triggered by an expanding HII region via radiatively driven implosion or collect-and-collapse. N117-3 is located within in the bubble, in projection. There is no obvious interpretation for this infall candidate's interaction with the associated bubble N117.

\citet{Myers1996} and \citet{Williams1999} present a model of infall that predicts line profiles similar to these observations. They assume two clouds (near and far) falling toward a common center and estimate the resulting line profiles accounting for optical depth effects as well as standard radial-dependencies of velocity and excitation temperature. \citet{Myers1996} show that an optically thick line and a higher excitation temperature in the cloud on the far side can produce a blue-shifted weighted line-shape. With further simplifications they show that by measuring five parameters, the Myers et al. (1996) model allows an estimate of the infall velocity. The measured parameters are: $\sigma$ (velocity dispersion of an optically thin tracer), T$_{BD}$ (the blue-shifted excess emission), T$_{RD}$ (the red-shifted emission), T$_D$ (the plateau emission), v$_{red}$ (the red-shifted peak emission velocity) and v$_{blue}$ (the blue-shifted peak emission velocity). See Figure 2 in \citet{Myers1996} for a diagram of these different quantities. When all quantities can be measured, the infall velocity is estimated to be:

\begin{equation*}
v_{in} \approx \frac{\sigma^2}{v_{red} - v_{blue}} \ln\left( \frac{1+e T_{BD}/T_D}{1+e T_{RD}/T_D}\right)
\end{equation*}

When the optical depth and V$_{in}$/$\sigma$ are sufficient large, the red peak can disappear (see Myers et al. 1996 for discussion of this effect). Thus, we are limited in our numerical analysis to N117-3. We estimated the relevant line parameters by eye. Our line profile measurements and infall velocity calculation are given in Table \ref{infalltable}. Since we do not have an optically thin measurement of this source, we have assumed the value of the velocity dispersion based on the optically thin $^{34}$CS observations by \citet{Williams1999}. They found a typical value to be 1.5 km/s. A smaller value would decrease the infall velocity (see equation above).

\begin{deluxetable}
{rr}
\label{infalltable}
\tablecaption{Infall Parameters}
\tablehead{}
\startdata
Object  &N117-3\\
T$_{BD}$  &0.9 K\\
T$_{RD}$  &0.2 K\\
T$_D$     &1.1 K\\
v$_{blue}$    &38.4 km/s\\
v$_{red}$     &40.9 km/s\\
$\sigma$    &1.5 km/s\\
v$_{in}$    &0.7 km/s\\
\enddata

\end{deluxetable}

Further analysis requires determining a distance. We assume a rotation curve following \citet{1993A&A...275...67B} and adopt the near kinematic distance of 4.4 kpc. We then can use the mid-IR integrated fluxes for toward N117-3 as measured by Herschel/Hi-Gal to estimate N117-3's mass and mass accretion rate. The mid-IR fluxes can be fit using the same modified blackbody model described above, yielding a mass of 96 M$_\odot$. We can further roughly estimate the mass infall rate \begin{math}
\dot M_{in} \end{math} using:

\begin{eqnarray*}
\dot M_{in} = 4 \pi R^2 v_{in} \rho\\
\rho = \frac{M}{4/3 \pi R^3}
\end{eqnarray*}

where R is the radius where infall has been detected, $v_{in}$ is the detected infall velocity and $\rho$ is the density of the infall gas. If we use the GBT beamsize projected to the near kinematic distance for R (0.32 pc), then we calculate a mass infall rate of 7 $\times$ 10$^{-5}$ M$_\odot$/yr. The dominant source of error in this calculation is likely due to the infall velocity. We estimate the uncertainty to be about a factor of 2. However, if we used a smaller value for R, as suggested by the small source size visible in the 8 $\mu$m GLIMPSE image, the mass infall rate would be proportionally smaller (by a factor of about 3). This result is consistent with massive or intermediate-mass star formation.

For the infall analysis, we have assumed an optically thick line. An alternative interpretation of these three line-profiles is that they are caused by alignment of two clouds along the line-of-sight. Observing an optically-thin tracer, such as $^{34}$CS would distinguish between these interpretations since the infall-model would predict a single-peak whereas the two cloud model predicts a double-peak.

N65-2, the other source which shows a non-Gaussian line shape, is stronger on the red-shifted side. This shape is not consistent with the infall model of Myers et al. (1996). This shape could be caused by two unrelated clouds along the line of the sight. There is further evidence of this interpretation in the map of N65 (see section 3.2).

\section{Conclusions}
We have surveyed 44 YSOs located near the edges of MIR-identified bubbles in CS(1-0) using the GBT. Our conclusions are:

\begin{itemize}
\item We have detected CS toward 18 sources.
\item Using Herschel/HiGal survey data, we calculated CS abundances for these sources to be $\sim$10$^{-7}$ and range between 0.16-9.36 $\times$10$^{-7}$.
\item Three sources show non-Gaussian line-profiles with strong emission on the blue-shifted side. We interpret this profile as caused by gas infall onto a protostar.
\item Two of the infall candidates (N62-1 and N90-2) are embedded in infrared dark clouds along the edge of their expanding bubbles. The combination of photometry-based YSO identification, CS-based infall, location inside an IRDC and on the edge of an expanding bubble is strongly suggestive of triggered star-formation.
\item Using a two-component model, we estimate that one infall candidate, N117-3, has an average infall speed of 0.31 km/s and a mass infall rate of 2.9 $\times$ 10$^{-5}$ M$_\odot$/yr. These numerical results are consistent with intermediate to massive star-formation.
\item Our interpretation of infall in N62-1, N90-2, and N117-3 assumes that the observed CS emission is optically thick.  However, our interpretation of the asymmetric, non-Gaussian line profile in N65-2 is that there are two line-of-sight clouds contributing to the emission.  It is possible that a similar mechanism could produce the profiles seen in N62-1, N90-2, and N117-3.  Further observations of an optically thin line, for example $^{34}$CS, are needed to distinguish between the two possible interpretations.
\end{itemize}
The three infall candidates are promising sources for further study to better determine the mechanisms involved in triggered star-formation. The two candidates embedded in IRDCs are especially promising and are being mapped in a follow-up study (Devine et al., in prep.).


\section{Acknowledgments}
The authors would like to acknowledge the constructive comments by an anonymous referee. David Frayer, of NRAO, assisted significantly during the observing and data reduction. CW and NQ were partially funded through Manchester University's Office of Academic Affairs.

\end{document}